\documentclass[aps,prl,twocolumn,showpacs,superscriptaddress,footinbib]{revtex4} 
\usepackage{graphicx}
\usepackage{dcolumn}
\usepackage{bm}
\usepackage{bbm}
\usepackage[utf8]{inputenc}
\usepackage{amssymb}
\usepackage{amsmath}
\usepackage{bbold}
\usepackage{pstricks}
\usepackage{slashed}
\usepackage{braket}
\usepackage{hyperref}
\usepackage{natbib}
\usepackage{float}
\usepackage{verbatim}
\usepackage{siunitx}
\usepackage{lipsum}
\usepackage{slashed}
\usepackage{ulem}

\definecolor{mygreen}{rgb}{0,0.5,0}
\definecolor{myblue}{rgb}{0,0,0.75}
\definecolor{mymagenta}{cmyk}{0,1,0,0.12}
\definecolor{mygray}{rgb}{0.5,0.5,0.5}

\newcommand{\gtext}[1]{{\color{mygreen} #1 }}

\newcommand{\Eq}[1]{Eq.~(\ref{#1})}

\newcommand{\Fig}[1]{Fig.~\ref{#1}}
\newcommand{\Figure}[1]{Figure~\ref{#1}}

\usepackage{umoline}

\newcommand{\newtext}[1]{\gtext{#1}}

\begin{document}

\title{Dynamical topological transitions in the massive Schwinger model with a $\theta$-term}

\author{T. V. Zache}
\email[]{zache@thphys.uni-heidelberg.de}
\affiliation{Heidelberg University, Institut f\"{u}r Theoretische Physik, Philosophenweg 16, 69120 Heidelberg, Germany}

\author{N. Mueller}
\affiliation{Physics Department, Brookhaven National Laboratory, Building 510A, Upton, New York 11973, USA}

\author{J. T. Schneider}
\affiliation{Heidelberg University, Institut f\"{u}r Theoretische Physik, Philosophenweg 16, 69120 Heidelberg, Germany}

\author{F. Jendrzejewski}
\affiliation{Heidelberg University, Kirchhoff-Institut f\"{u}r Physik,
Im Neuenheimer Feld 227, 69120 Heidelberg, Germany}

\author{J. Berges}
\affiliation{Heidelberg University, Institut f\"{u}r Theoretische Physik, Philosophenweg 16,
69120 Heidelberg, Germany}

\author{P. Hauke}
\affiliation{Heidelberg University, Institut f\"{u}r Theoretische Physik, Philosophenweg 16,
69120 Heidelberg, Germany}
\affiliation{Heidelberg University, Kirchhoff-Institut f\"{u}r Physik,
Im Neuenheimer Feld 227, 69120 Heidelberg, Germany}

\date{\today}

\begin{abstract}
Aiming at a better understanding of anomalous and topological effects in gauge theories out-of-equilibrium, 
we study the real-time dynamics of a prototype model for CP-violation, the massive Schwinger model with a $\theta$-term. 
We identify dynamical quantum phase transitions between different topological sectors that appear after sufficiently strong quenches of the $\theta$-parameter. Moreover, we establish a general dynamical topological order parameter, which can be accessed through fermion two-point correlators and, importantly, which can be applied for interacting theories. 
Enabled by this result, we show that the topological transitions persist beyond the weak-coupling regime. 
Finally, these effects can be observed with table-top experiments based on existing cold-atom, superconducting-qubit, and trapped-ion technology. 
Our work, thus, presents a significant step towards quantum simulating topological and anomalous real-time phenomena relevant to nuclear and high-energy physics.
\end{abstract}
\maketitle

{\it Introduction.}  
The topological structure of gauge theories has many important manifestations~\cite{Klinkhamer:1984di,Dashen:1974ck,Soni:1980ps,Boguta:1983xs,Forgacs:1983yu}.  In quantum chromodynamics (QCD), e.g., it allows for an additional term in the action that explicitly breaks charge conjugation parity ($CP$) symmetry \cite{tHooft:1976snw,Jackiw:1976pf,Callan:1979bg}. 
Though the angle $\theta$ that parametrizes this term is in principle unconstrained, experiments have found very strong bounds on $CP$ violation, consistent with $\theta=0$~\cite{chupp2017electric}. 
In one elegant explanation, $\theta$ is described as a dynamical field that undergoes a phase transition, the `axion' \cite{Weinberg:1977ma,Wilczek:1977pj,Peccei:1977hh}, which is currently sought after in experiments \cite{Graham:2015ouw}. 
However, the controlled study of topological effects far from equilibrium remains highly challenging~\cite{Mace:2016svc}. 
So-called quantum simulators offer an attractive alternative approach. 
These are engineered quantum devices that mimic desired Hamiltonians in an analog way or synthesize them on digital (qubit based) quantum computers \cite{cirac2012goals,hauke2012can,nuclearReview}. 
While theories of the standard model, such as QCD, are beyond the current abilities of quantum simulators, 
existing technology \cite{martinez2016real,klco2018quantum} can already simulate simpler models that put insights into the topological properties of gauge theories within reach. 
In this respect, the massive Schwinger model~\cite{Coleman:1976uz}, describing quantum electrodynamics (QED) in 1+1 dimensions, is particularly interesting because it allows for a $CP$-odd $\theta$-term similar to QCD. 
However, while ground state and thermal properties of QCD and of the Schwinger model have been extensively studied \cite{coleman1975charge,Petreczky:2012rq}, much less is known about their topological structure out of equilibrium. 

In this work, we investigate the non-equilibrium real-time evolution of the massive Schwinger model after a quench of the topological $\theta$ angle. 
We find topological transitions in the fermion sector, which appear as vortices in the single-particle propagator when $\theta$ changes by more than a critical value. 
In the limit of vanishing gauge coupling, we rigorously connect this phenomenon to dynamical quantum phase transitions (DQPTs), which in condensed-matter lattice models are currently receiving considerable attention \cite{heyl2013dynamical,flaschner2016observation,jurcevic2017direct,heyl2018dynamical}. 
A topological nature of DQPTs has previously been revealed in non-interacting theories \cite{budich2016dynamical,tian2018direct,xu2018measuring}. 
Here, we demonstrate how to construct a general dynamical topological invariant that is valid in the continuum and, most importantly, also in interacting theories. 
Moreover, our topological invariant provides a physical interpretation of DQPTs in terms of fermionic correlation functions. 
Enabled by this result, we use non-perturbative real-time lattice calculations at intermediate to strong coupling to show that the topological transition persists up to $\frac{e}{m}\lesssim 1$. 
Already for lattices as small as 8 sites, we obtain good infrared convergence. 
Moreover, the relevant phenomena occur on time scales that have already been accessed in proof-of-principle quantum simulations of gauge theories ~\cite{martinez2016real,klco2018quantum}. 
These features will enable near-future experiments based on trapped ions~\cite{martinez2016real}, superconducting qubits~\cite{klco2018quantum} and cold neutral atoms~\cite{zache2018quantum} to probe this dynamical topological transition.

{\it $\theta$-quenches in the massive Schwinger model.} The massive Schwinger model is a prototype model for 3+1D QCD since both share important features such as a non-trivial topological vacuum structure and a chiral anomaly \cite{coleman1975charge,Coleman:1976uz}. 
$CP$ violation can be studied by adding a so-called topological $\theta$-term, $\frac{e\theta}{2\pi}E_x$, to the Hamiltonian density, where $E$ is the electric field and $e$ the dimensionful gauge coupling. 
In temporal axial gauge, and by making a chiral transformation, the $\theta$-term can be absorbed into the fermion mass term to give the following Hamiltonian~\cite{Coleman:1976uz}, 
\begin{align}\label{eq:Hamiltonianbasis}
{H}_\theta = \int dx \, \left[  \frac{1}{2}{E}^2_x + {\psi}_x^\dagger \gamma^0 \left(i \gamma^1 D_x + m\, 
\text{e}^{i \theta \gamma^5}\right) {\psi}_x\right] \; .
\end{align}
Here, ${\psi}$ are two-component fermion operators, $m$ the fermion rest mass, $\gamma^{0/1}$ constitute a two-dimensional Clifford algebra, and $\gamma^5 \equiv \gamma^0 \gamma^1$. 
The first term describes the energy of the electric field, which is coupled to the kinetic energy of the fermionic matter via the covariant derivative ${D_x}$. 

Here, we wish to study how topological properties appearing through the $CP$-violating $\theta$ term become manifest in the real-time dynamics of the theory. 
To this end, we prepare the system in the ground-state $|\Omega(\theta) \rangle$ of ${H}_\theta$ and switch abruptly to another value $\theta'$, thereby quenching the system out of equilibrium. 
Since the $\theta$-angle in the massive Schwinger model has the same topological origin as its counterpart in 3+1D QCD, we can interpret the studied quench as a classical, external axion field.
In the following, we will show that this quench generates topological transitions, which appear as momentum--time vortices of the phase of the gauge-invariant time-ordered Green's function,
\begin{align}\label{eq:defphase}
g_{\theta \rightarrow \theta'}(k,t) = \int dx \, e^{-ikx} \langle \psi^\dagger(x,t) e^{-ie \int^x_0 dx' \, A(x',t)} \psi(0,0) \rangle .
\end{align}
We first discuss these topological transitions in the continuum theory at weak coupling, where we show analytically their direct correspondence to DQPTs. 
The weak-coupling results will motivate the definition of a general topological invariant, which will enable us to study also the interacting theory, discussed further below.

\begin{figure}
	\centering{
		\includegraphics[scale=0.25]{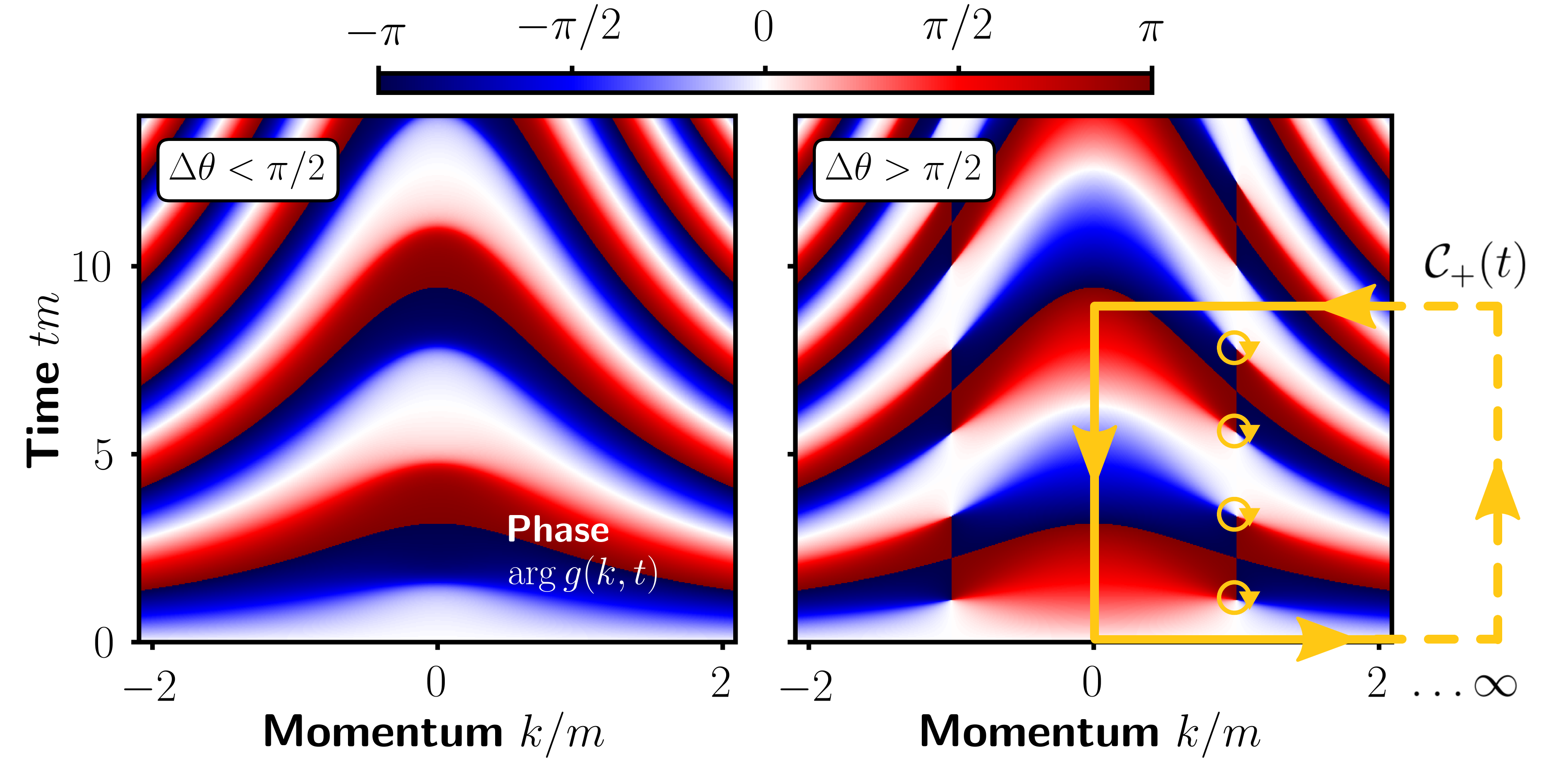}\caption{\label{fig:1} 
			Phase of the time-ordered correlator [\Eq{eq:defphase}] after $\theta$ quenches at vanishing gauge coupling. 
			The real-time evolution of the phase exhibits qualitative differences when the quench is weaker/stronger than the critical value $\Delta\theta_c = \pi/2$, exemplified here for $\Delta\theta=0.45 \pi$ (left) and $\Delta\theta=\pi$ (right). While for small quenches $|\Delta\theta|<\Delta\theta_c$ the phase is analytic, for large quenches $|\Delta\theta|>\Delta\theta_c$ vortices form at $(\pm k_c,t_c^{(n)})$. 
			The integration path $\mathcal{C}_+(t)$, here shown for $tm \approx 9$, encloses a discrete number of vortices (marked by yellow circles), leading to
			integer increments of the topological invariant $\nu$ as time progresses (see \Fig{fig:2}).}}
\end{figure} 

\textit{Weak-coupling limit.} 
In the weak-coupling limit, $e/m\to 0$, the massive Schwinger model is a free fermionic theory that can be solved analytically by diagonalizing $H_{\theta}=\int dk H_{\theta}(k)$, with $H_{\theta}(k) = {\psi}_k^\dagger \gamma^0 \left(k \gamma^1  + m\, \text{e}^{i \theta \gamma^5}\right) {\psi}_k$.  
\Figure{fig:1} displays the phase of  $g_{\theta \rightarrow \theta'}$ as a function of $(k,t)$ for two exemplary quenches with $\Delta \theta = 0.45\pi, \pi$ 
(our results here depend only on $\Delta\theta=(\theta-\theta') \in (-\pi,\pi]$). 
Strong quenches in the range $|\Delta\theta| > \frac{\pi}{2}$ are accompanied by the formation of vortices at critical times $t_c^{(n)}=(2n-1)t_c$, with $t_c=\frac{\pi}{2\omega(k_c)}$, $n \in \mathbb{N}$ and $\omega(k)=\sqrt{k^2+m^2}$.  
These appear in pairs of opposite winding at critical modes $\pm k_c=\pm m \sqrt{-\cos\left(\Delta \theta\right)}$. 

\begin{figure}
	\centering{\includegraphics[scale=0.3]{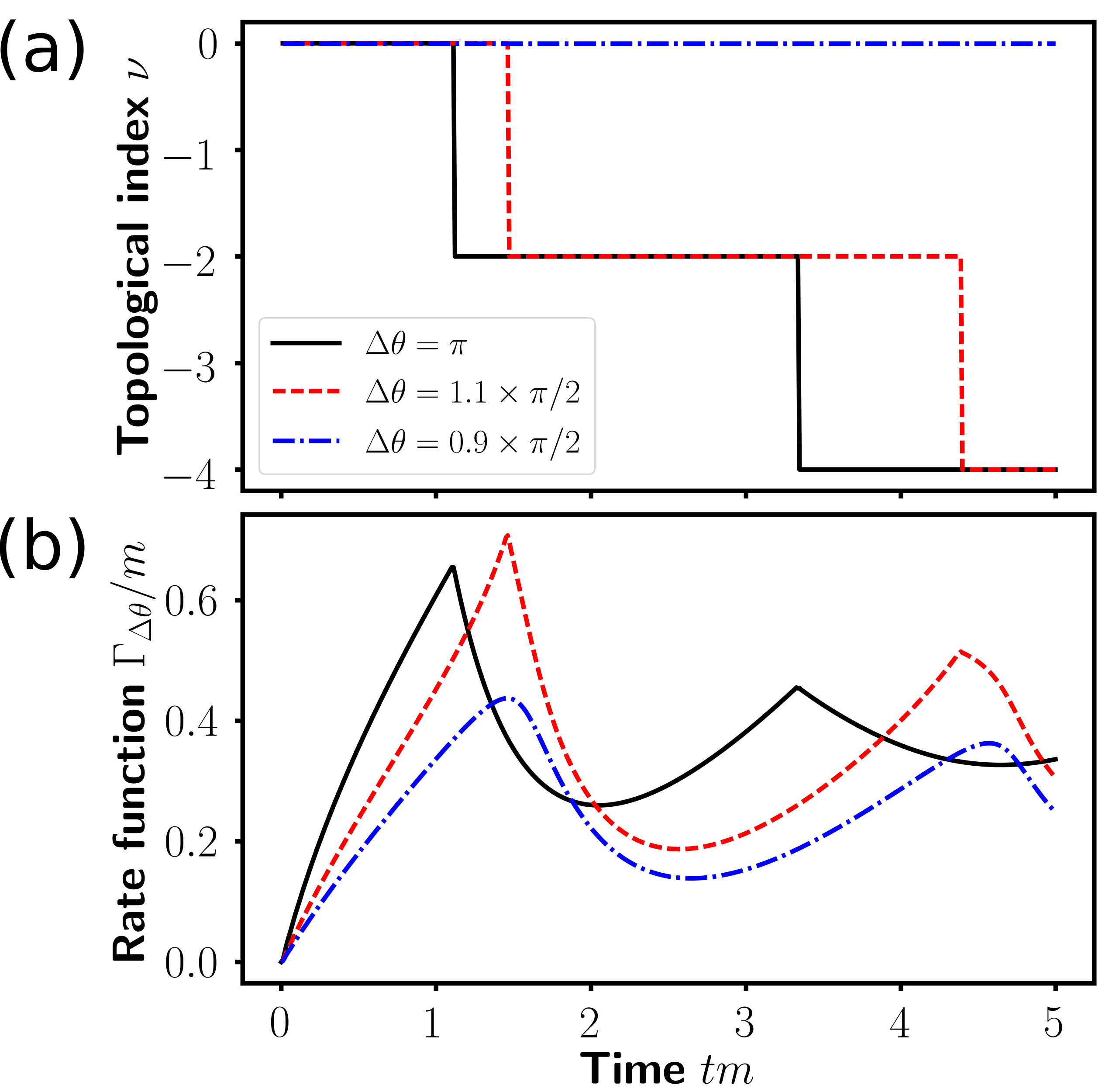}
		\caption{\label{fig:2}
				Dynamical topological transitions at vanishing gauge coupling. 
				(a) The topological invariant exhibits jumps at critical times $t_c^{(n)} = (2n-1)\pi/\left[2\omega(k_c)\right]$ with $n\in \mathbb{N}$, if $|\Delta\theta| > {\pi}/{2}$, while the dynamics is topologically trivial for $\left|\Delta\theta\right| <  {\pi}/{2}$. 
				(b) For $|\Delta\theta| > {\pi}/{2}$, the rate function [\Eq{eq:ratefunction}] shows non-analytic kinks at times $t_c^{(n)}$. 
		}
	}
\end{figure}

This observation suggests to define a dynamical topological order parameter that counts the difference of vortices contained in left ($-$) versus right ($+$) moving modes, $\nu\equiv n_+-n_-$, with
\begin{align}\label{eq:rel_topInd}
n_\pm (t) \equiv \frac{1}{2\pi} \oint_{\mathcal{C}_\pm(t)} d\mathbf{z} \, \left\{ \tilde{g}^\dagger(\mathbf{z}) \nabla_\mathbf{z} \tilde{g}(\mathbf{z})\right\} \; .
\end{align}
Here, $\tilde{g}(\mathbf{z})\equiv g_{\theta \rightarrow \theta'}(k,t')/ |g_{\theta \rightarrow \theta'}(k,t')|$ and $\mathcal{C}_\pm(t)$ is a rectangular path enclosing the left/right half of the $\mathbf{z} =(k,t')$-plane up to the present time $t$, i.e., it runs 
(counter-clockwise) along $(0,0)\leftrightarrow (0,t)\leftrightarrow(\pm\infty,t)\leftrightarrow(\pm\infty,0)\leftrightarrow(0,0)$ as visualized in \Fig{fig:1}. 
As exemplified in \Fig{fig:2}(a), the topological invariant remains trivial for $|\Delta \theta|<\pi/2$, while for $|\Delta \theta|>\pi/2$ it changes abruptly at critical times $t_c^{(n)}$. 

These singular times coincide with fundamental changes in the properties of the real-time evolution, coined DQPTs \cite{heyl2013dynamical}. 
DQPTs are revealed in the so-called Loschmidt amplitude, which is related to the vacuum persistence amplitude~\cite{Gelis:2015kya} and which is a common measure, e.g., in the field of quantum chaos~\cite{gorin2006dynamics}. 
The Loschmidt amplitude quantifies the overlap of the time-evolved state with its initial condition, 
\begin{align}\label{eq:Loschmidt}
L_{\theta \rightarrow \theta'}(t) \equiv \langle \Omega(\theta) | \mathrm{e}^{-iH_{\theta'}t} | \Omega(\theta) \rangle \; .
\end{align}
It is convenient to further define an intensive `rate function' 
\begin{align}\label{eq:ratefunction}
\Gamma_{\theta \rightarrow \theta'}(t) \equiv -\lim\limits_{V \rightarrow \infty}\text{Re} \left\lbrace \frac{1}{V}\log \left[L_{\theta \rightarrow \theta'}(t)\right] \right\rbrace \,.
\end{align} 
DQPTs appear as non-analyticities of \Eq{eq:ratefunction} [zeros of \Eq{eq:Loschmidt}]. 

In the limit $e/m\to 0$, the system is in a product state $| \Omega(\theta) \rangle = \bigotimes_k | \Omega_k(\theta) \rangle$. The Loschmidt amplitude can then be decomposed into Fourier modes, 
\begin{equation}
\label{eq:Loschmidt_modes}
L_{\theta \rightarrow \theta'}(t) = \prod_k \langle \Omega_k(\theta)| \mathrm{e}^{-iH_{\theta'}(k)t}| \Omega_k(\theta) \rangle\,,
\end{equation} 
Since at $e/m\to 0$ we have the additional identity $\langle \Omega_k(\theta)| \mathrm{e}^{-iH_{\theta'}(k)t}| \Omega_k(\theta)\rangle = g_{\theta \rightarrow \theta'}(k,t)$, zeros of the Loschmidt amplitude imply that the phase of the Green's function becomes undefined for a critical mode, enabling the appearance of the vortices seen in \Fig{fig:1}. 
As a consequence, at zero coupling the topological transitions and non-analyticities of the rate function in \Eq{eq:ratefunction} strictly coincide [see \Fig{fig:2}(b)].  

For non-interacting lattice theories, a topological nature of DQPTs has previously been revealed through the phase of the Fourier-decomposed Loschmidt amplitude, 
$\arg\left[\langle \Omega_k(\theta)| \exp [-iH_{\theta'}(k)t]| \Omega_k(\theta) \rangle\right]=\phi_{\mathrm{geom}}+\phi_{\mathrm{dyn}}$ ~\cite{budich2016dynamical}.  
Here, the total phase has been divided into a trivial dynamical phase $\phi_{\mathrm{dyn}}(k,t)$ and the so-called Pancharatnam geometric phase, $\phi_{\mathrm{geom}}(k,t)$. 
At a DQPT, the winding number of $\phi_{\mathrm{geom}}$ acquires a singular change. 
This change can be computed by integration across (half) the Brillouin zone at fixed time $t$~\cite{budich2016dynamical}, which has been used in the recent experiments of Refs.~\cite{tian2018direct,xu2018measuring}. 
For this prescription to work, however, one needs to subtract the trivial dynamical phase $\phi_{\mathrm{dyn}}$, which can reasonably be obtained only perturbatively close to the non-interacting case. 
Compared to this standard prescription, our construction in \Eq{eq:rel_topInd} has a number of advantages.  
First, the prescription of Ref.~\cite{budich2016dynamical} fails for $\theta \neq 0, \pi$, where the absence of a particle--hole symmetry makes modes at $k=0,\pm\infty$ inequivalent. 
Second, and more importantly, by using a closed path in the $(k,t)$ plane (cf.\ \Fig{fig:1}) only the singular geometric part contributes to the integral in \Eq{eq:rel_topInd}, irrespective of the smooth dynamical phase. 
Thus, together with the definition through fermionic correlators, \Eq{eq:defphase}, instead of Fourier modes of the wave-function overlap, \Eq{eq:Loschmidt_modes}, our formulation enables us to tackle also the interacting theory.

\begin{figure*}
\centering{\includegraphics[scale=0.35
]{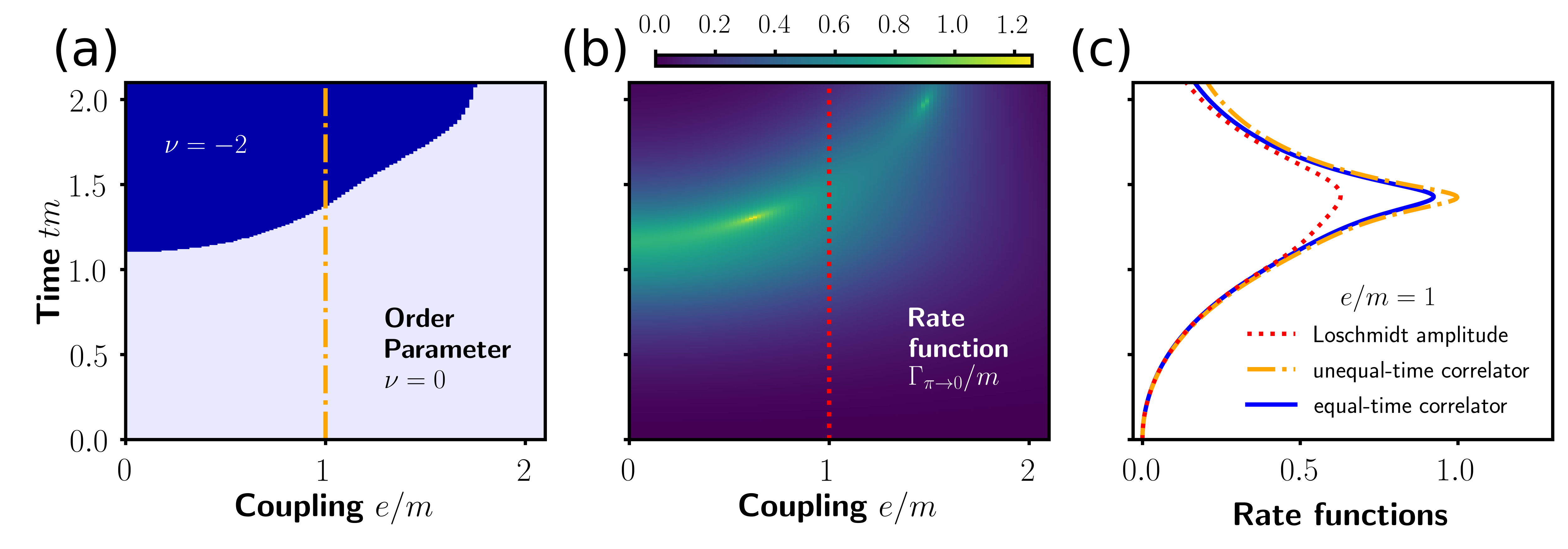}
\caption{\label{fig:Loschmidt_interacting}
	Dynamical topological transitions beyond weak coupling. 
	$(a)$ The integer-valued topological invariant $\nu$ clearly distinguishes different `phases' in the $(t,e)$-plane. The topological transition persists at larger coupling, but shifts towards later times and appears to vanish for $e/m \gtrsim 2$. 
	$(b)$ The maxima of the rate function obtained from the many-body overlap agree qualitatively with the transitions in $\nu$, but are blurred by the finite lattice size.  
	$(c)$ 
	Rate functions computed from the full wave-function overlap [red dotted; c.f. panel $(b)$ and \Eq{eq:ratefunction}], from fermionic two-time correlators [orange dot-dashed; c.f. panel $(a)$ and \Eq{eq:defphase}], and equal-time correlators [blue solid line; c.f. \Eq{eq:absvalue}], all indicate the same time of the first topological transition. 
	Simulations are for a small lattice of $N = 8$ sites as relevant for first quantum-simulator experiments, and with lattice spacing $am = 0.8$. 
}}
\end{figure*}

{\it Towards strong coupling.}
To investigate if the topological transitions persist at non-vanishing coupling, $e/m>0$, we perform non-perturbative real-time lattice simulations based on Exact Diagonalization (ED). We focus on the strongest quench $\Delta\theta = \pi$ (or $-m \rightarrow m$), using staggered fermions with lattice Hamiltonian \cite{Banks:1975gq}
\begin{align}\label{eq:Hamiltonian_staggered}
{H} &= a\sum\limits_{{n=0}}^{N-1} \left[\frac{{E}_{n}^2}{2} +  m\left(-1\right)^{{n}}  {\phi}_{n}^\dagger  {\phi}_{n}  -\frac{i}{2a}  \left({\phi}_{n}^\dagger {U}_{n} {\phi}_{{n}+1} - \text{h.c.}\right) \right] .
\end{align}
Here, $\phi_n$ are one-component fermion operators on an even number of lattice sites $N$, ${E}_n$ and ${U}_n$ are electric fields and links, and $a$ is the lattice spacing. To apply ED, we restrict the simulation to the physical Hilbert space by solving the Gauß law constraint $G_n|\text{phys} \rangle = 0$ with $G_n = E_n - E_{n-1} - e\left[\phi_n^\dagger  \phi_n + \frac{\left(-1\right)^{n}-1}{2}\right]$.  
In contrast to previous works \cite{hamer1997series,martinez2016real}, we use periodic boundary conditions (PBC) \footnote{To obtain a finite-dimensional Hilbert space, we drop the single remaining bosonic mode describing a constant background field.}, see \cite{longpaper} for more details. 
To efficiently compute the topological invariant $\nu$, we adapt a formalism that has originally been developed for computing Chern numbers in momentum space~\cite{fukui2005chern}. 
The possibility to adapt this formalism to our case is another feature of our definition in \Eq{eq:rel_topInd} since it is enabled by the use of a closed integration path in the $(k,t)$ plane. 
This adaption forces $\nu$ to remain integer-valued even when evaluated on coarse grids, thus leading to convergence already for small lattices \cite{longpaper}. 

As can be expected from the above discussions, at small $e/m$ transitions in the topological invariant coincide with maxima in the rate function, see \Fig{fig:Loschmidt_interacting}. 
Further, both structures smoothly connect to larger values of $e/m$. 
Importantly, however, while the system sizes accessible for ED do not allow one to discern clear kinks in the rate function, the non-equilibrium topological invariant $\nu$ sharply distinguishes between topologically inequivalent phases, revealing a shift of the transitions towards larger $t_c$ as $e/m$ is increased.  
While the results for $e/m\lesssim 1$ are already reasonably finite-volume converged for the small system size plotted, at $e/m\gtrsim 1$ finite-volume effects persist up to $N=20$ (not shown here; c.f.\ \cite{longpaper}).  
Nevertheless, the topological transition seems to vanish altogether at sufficiently large coupling $e/m \gtrsim 2$, in agreement with 
$\theta$ being an irrelevant parameter in the limit $m \rightarrow 0$~\cite{abdalla1991non}.

{\it Quantum simulation.} 
Importantly, the first topological transition happens on times of order $t_c m\sim 1-2$, which lies within coherence times that are achievable with existing and proposed quantum simulators \cite{martinez2016real,klco2018quantum,zache2018quantum}. 
A straightforward realization of the scenario discussed in this letter may be achieved with a quantum computer based on trapped-ions or superconducting qubits, where quench dynamics has been studied recently~\cite{martinez2016real,klco2018quantum}. 
Though these experiments used only four lattice sites of staggered fermions, larger lattices are within reach of current technology \cite{monz2016realization,barends2016digitized,kandala2017hardware,landsman2018verified}. Alternatively, various works have proposed analogue quantum simulators of the massive Schwinger model \cite{wiese2013ultracold,zohar2015quantum,dalmonte2016lattice,Magnifico:2018wek}. One possible implementation is based on a mixture of bosonic and fermionic atoms in a tilted optical lattice~\cite{zache2018quantum}, where the fermion mass corresponds to Rabi oscillations between two hyperfine states driven by radiofrequency radiation. In this setup, a mass quench may be simply implemented by abruptly adjusting the corresponding Rabi frequency.

These experiments may unveil the topological transitions through different observables: First, a digital quantum computer could in principle work with the many-body wavefunctions to directly calculate the order parameter $\nu$ [Eq. \eqref{eq:rel_topInd}] and the rate function $\Gamma_{\theta \rightarrow \theta'}(t)$ [Eq. \eqref{eq:ratefunction}]. Second, one could measure the two-time correlator $g_{\theta \rightarrow \theta'}(k,t)$ [Eq. \eqref{eq:defphase}]~\cite{knap2013probing,uhrich2018probing} and thereby avoid the study of many-body overlaps. 
Third, the discrete transition points of the order parameter are indicated also in experimentally more accessible equal-time correlation functions, $[\underline{F}(t)]_{xy}^{\alpha \beta}\equiv\left\langle \left[\psi^{\alpha}(t,x), \bar{\psi}^{\beta}(t,y) \right] \right\rangle$.
Namely, let us define 
\begin{align}\label{eq:absvalue}
K_{\theta\rightarrow \theta'} (t) &\equiv \prod_k \left[\mathbf{F}(k,t) + \mathbf{F}(k,0)\right]^2\,,
\end{align}
where $\mathbf{F}=(F_{s},F_{1},F_{5})$ are the Lorentz components of the correlator, $\underline{F}(t) = F_{s}(t) \mathbf{1} + F_{\mu}(t) \gamma^\mu + i F_{5}(t) \gamma^5$. 
In the weak-coupling limit, one has $K_{\theta\rightarrow \theta'} (t) = \prod_k |g_{\theta \rightarrow \theta'}(k,t)|^2= |L_{\theta \rightarrow \theta'}(t)|^2$ (for details, see \cite{longpaper}). 
We thus have three complementary definitions that coincide for $e/m \rightarrow 0$, obtained from equal-time correlators, \Eq{eq:absvalue}, two-time correlators, \Eq{eq:defphase}, and the full many-body Loschmidt amplitude, \Eq{eq:Loschmidt}. 
Remarkably, even at intermediate couplings the rate functions from all three indicate the same critical times of the dynamical quantum phase transition, shown in \Fig{fig:Loschmidt_interacting}(c) for $e/m=1$. 

Besides its experimental simplicity, \Eq{eq:absvalue} also gives an interesting interpretation of the dynamical topological transition in terms of a dephasing effect. 
Namely, \Eq{eq:absvalue} has singularities if and only if the mode $k_c$ at time $t_c$ exhibits perfect anti-correlation with the initial state, $\mathbf{F}(k_c,t_c)=-\mathbf{F}(k_c,0)$. 
This anti-correlation may be interpreted as the time evolved $\mathbf{F}(k_c,t_c)$ being the chiral transform of the initial $\mathbf{F}(k_c,0)$ with transformation parameter $\pi/2$.

\textit{Conclusions. }
In this manuscript, we have studied the real-time dynamics of massive 1+1D QED with a $\theta$-term, as a prototype model for topological effects in gauge theories. 
By establishing a general dynamical topological order parameter, which can be obtained from fermionic correlators and is valid in interacting theories, we have identified the appearance of dynamical topological transitions after changes in the external `axion' field.  
A connection between the topological transitions to DQPTs, which is rigorous at zero coupling, persists in our numerics of the interacting theory, thus providing a physical interpretation of DQPTs in terms of fermionic correlators. 
Finally, our topological order parameter can directly be applied also in the study of condensed-matter models, where the construction of topological invariants for interacting systems is a major outstanding challenge \cite{gurarie2011single,rachel2018interacting}. 

In our study, we have identified a relevant problem for state-of-the-art quantum simulation.  The described dynamical transitions constitute an ideal first step, because the relevant dynamics appears at short time scales and small system sizes. We expect the topological nature to provide robustness against experimental imperfections, which may provide a starting point to tackle the question of certifiability of quantum simulation. 

Despite the simplicity of the considered model, our study shows that quantum simulators provide a unique perspective to the topological structure of QCD out of equilibrium.
Phenomena closely related to physics studied in this article are the conjectured Chiral Magnetic and similar effects~\cite{Kharzeev:2007jp,Fukushima:2008xe,Kharzeev:2015znc,Skokov:2016yrj}, which remain challenging in and out of equilibrium for theoretical studies~\cite{Son:2009tf,Yee:2009vw,Son:2012wh,Stephanov:2012ki,Chen:2013iga,Mace:2016svc, Mueller:2016ven}. Here, a simple next step for future quantum simulation is to model these effects by spatial domains of the $\theta$-parameter \cite{Tuchin:2018rrw}.

{\it Note added.} 
For a related work on dynamical quantum phase transitions in lattice gauge theories, see the article published on the arxiv on the same day by Yi-Ping Huang, Debasish Banerjee, and Markus Heyl.

{\it Acknowledgments.} 
This work is part of and supported by the DFG Collaborative Research Centre ``SFB 1225 (ISOQUANT)'', the ERC Advanced Grant ``EntangleGen'' (Project-ID 694561), and
the Excellence Initiative of the German federal government and the state governments – funding line Institutional Strategy (Zukunftskonzept): DFG project number ZUK 49/Ü. NM is supported by the U.S. Department of Energy, Office of Science, Office of Nuclear Physics, under contract No. DE- SC0012704. 
\bibliographystyle{apsrev4-1} 
\bibliography{references}

\begin{thebibliography}{60}%
\makeatletter
\providecommand \@ifxundefined [1]{%
 \@ifx{#1\undefined}
}%
\providecommand \@ifnum [1]{%
 \ifnum #1\expandafter \@firstoftwo
 \else \expandafter \@secondoftwo
 \fi
}%
\providecommand \@ifx [1]{%
 \ifx #1\expandafter \@firstoftwo
 \else \expandafter \@secondoftwo
 \fi
}%
\providecommand \natexlab [1]{#1}%
\providecommand \enquote  [1]{``#1''}%
\providecommand \bibnamefont  [1]{#1}%
\providecommand \bibfnamefont [1]{#1}%
\providecommand \citenamefont [1]{#1}%
\providecommand \href@noop [0]{\@secondoftwo}%
\providecommand \href [0]{\begingroup \@sanitize@url \@href}%
\providecommand \@href[1]{\@@startlink{#1}\@@href}%
\providecommand \@@href[1]{\endgroup#1\@@endlink}%
\providecommand \@sanitize@url [0]{\catcode `\\12\catcode `\$12\catcode
  `\&12\catcode `\#12\catcode `\^12\catcode `\_12\catcode `\%12\relax}%
\providecommand \@@startlink[1]{}%
\providecommand \@@endlink[0]{}%
\providecommand \url  [0]{\begingroup\@sanitize@url \@url }%
\providecommand \@url [1]{\endgroup\@href {#1}{\urlprefix }}%
\providecommand \urlprefix  [0]{URL }%
\providecommand \Eprint [0]{\href }%
\providecommand \doibase [0]{http://dx.doi.org/}%
\providecommand \selectlanguage [0]{\@gobble}%
\providecommand \bibinfo  [0]{\@secondoftwo}%
\providecommand \bibfield  [0]{\@secondoftwo}%
\providecommand \translation [1]{[#1]}%
\providecommand \BibitemOpen [0]{}%
\providecommand \bibitemStop [0]{}%
\providecommand \bibitemNoStop [0]{.\EOS\space}%
\providecommand \EOS [0]{\spacefactor3000\relax}%
\providecommand \BibitemShut  [1]{\csname bibitem#1\endcsname}%
\let\auto@bib@innerbib\@empty
\bibitem [{\citenamefont {Klinkhamer}\ and\ \citenamefont
  {Manton}(1984)}]{Klinkhamer:1984di}%
  \BibitemOpen
  \bibfield  {author} {\bibinfo {author} {\bibfnamefont {F.~R.}\ \bibnamefont
  {Klinkhamer}}\ and\ \bibinfo {author} {\bibfnamefont {N.~S.}\ \bibnamefont
  {Manton}},\ }\href {\doibase 10.1103/PhysRevD.30.2212} {\bibfield  {journal}
  {\bibinfo  {journal} {Phys. Rev.}\ }\textbf {\bibinfo {volume} {D30}},\
  \bibinfo {pages} {2212} (\bibinfo {year} {1984})}\BibitemShut {NoStop}%
\bibitem [{\citenamefont {Dashen}\ \emph {et~al.}(1974)\citenamefont {Dashen},
  \citenamefont {Hasslacher},\ and\ \citenamefont {Neveu}}]{Dashen:1974ck}%
  \BibitemOpen
  \bibfield  {author} {\bibinfo {author} {\bibfnamefont {R.~F.}\ \bibnamefont
  {Dashen}}, \bibinfo {author} {\bibfnamefont {B.}~\bibnamefont {Hasslacher}},
  \ and\ \bibinfo {author} {\bibfnamefont {A.}~\bibnamefont {Neveu}},\ }\href
  {\doibase 10.1103/PhysRevD.10.4138} {\bibfield  {journal} {\bibinfo
  {journal} {Phys. Rev.}\ }\textbf {\bibinfo {volume} {D10}},\ \bibinfo {pages}
  {4138} (\bibinfo {year} {1974})}\BibitemShut {NoStop}%
\bibitem [{\citenamefont {Soni}(1980)}]{Soni:1980ps}%
  \BibitemOpen
  \bibfield  {author} {\bibinfo {author} {\bibfnamefont {V.}~\bibnamefont
  {Soni}},\ }\href {\doibase 10.1016/0370-2693(80)90104-5} {\bibfield
  {journal} {\bibinfo  {journal} {Phys. Lett.}\ }\textbf {\bibinfo {volume}
  {93B}},\ \bibinfo {pages} {101} (\bibinfo {year} {1980})}\BibitemShut
  {NoStop}%
\bibitem [{\citenamefont {Boguta}(1983)}]{Boguta:1983xs}%
  \BibitemOpen
  \bibfield  {author} {\bibinfo {author} {\bibfnamefont {J.}~\bibnamefont
  {Boguta}},\ }\href {\doibase 10.1103/PhysRevLett.50.148} {\bibfield
  {journal} {\bibinfo  {journal} {Phys. Rev. Lett.}\ }\textbf {\bibinfo
  {volume} {50}},\ \bibinfo {pages} {148} (\bibinfo {year} {1983})}\BibitemShut
  {NoStop}%
\bibitem [{\citenamefont {Forgacs}\ and\ \citenamefont
  {Horvath}(1984)}]{Forgacs:1983yu}%
  \BibitemOpen
  \bibfield  {author} {\bibinfo {author} {\bibfnamefont {P.}~\bibnamefont
  {Forgacs}}\ and\ \bibinfo {author} {\bibfnamefont {Z.}~\bibnamefont
  {Horvath}},\ }\href {\doibase 10.1016/0370-2693(84)91926-9} {\bibfield
  {journal} {\bibinfo  {journal} {Phys. Lett.}\ }\textbf {\bibinfo {volume}
  {138B}},\ \bibinfo {pages} {397} (\bibinfo {year} {1984})}\BibitemShut
  {NoStop}%
\bibitem [{\citenamefont {'t~Hooft}(1976)}]{tHooft:1976snw}%
  \BibitemOpen
  \bibfield  {author} {\bibinfo {author} {\bibfnamefont {G.}~\bibnamefont
  {'t~Hooft}},\ }\href {\doibase 10.1103/PhysRevD.18.2199.3,
  10.1103/PhysRevD.14.3432} {\bibfield  {journal} {\bibinfo  {journal} {Phys.
  Rev.}\ }\textbf {\bibinfo {volume} {D14}},\ \bibinfo {pages} {3432} (\bibinfo
  {year} {1976})},\ \bibinfo {note} {[,70(1976)]}\BibitemShut {NoStop}%
\bibitem [{\citenamefont {Jackiw}\ and\ \citenamefont
  {Rebbi}(1976)}]{Jackiw:1976pf}%
  \BibitemOpen
  \bibfield  {author} {\bibinfo {author} {\bibfnamefont {R.}~\bibnamefont
  {Jackiw}}\ and\ \bibinfo {author} {\bibfnamefont {C.}~\bibnamefont {Rebbi}},\
  }\href {\doibase 10.1103/PhysRevLett.37.172} {\bibfield  {journal} {\bibinfo
  {journal} {Phys. Rev. Lett.}\ }\textbf {\bibinfo {volume} {37}},\ \bibinfo
  {pages} {172} (\bibinfo {year} {1976})},\ \bibinfo {note}
  {[,353(1976)]}\BibitemShut {NoStop}%
\bibitem [{\citenamefont {Callan}\ \emph {et~al.}(1979)\citenamefont {Callan},
  \citenamefont {Dashen},\ and\ \citenamefont {Gross}}]{Callan:1979bg}%
  \BibitemOpen
  \bibfield  {author} {\bibinfo {author} {\bibfnamefont {C.~G.}\ \bibnamefont
  {Callan}, \bibfnamefont {Jr.}}, \bibinfo {author} {\bibfnamefont {R.~F.}\
  \bibnamefont {Dashen}}, \ and\ \bibinfo {author} {\bibfnamefont {D.~J.}\
  \bibnamefont {Gross}},\ }\href {\doibase 10.1103/PhysRevD.20.3279} {\bibfield
   {journal} {\bibinfo  {journal} {Phys. Rev.}\ }\textbf {\bibinfo {volume}
  {D20}},\ \bibinfo {pages} {3279} (\bibinfo {year} {1979})}\BibitemShut
  {NoStop}%
\bibitem [{\citenamefont {Chupp}\ \emph {et~al.}(2017)\citenamefont {Chupp},
  \citenamefont {Fierlinger}, \citenamefont {Ramsey-Musolf},\ and\
  \citenamefont {Singh}}]{chupp2017electric}%
  \BibitemOpen
  \bibfield  {author} {\bibinfo {author} {\bibfnamefont {T.}~\bibnamefont
  {Chupp}}, \bibinfo {author} {\bibfnamefont {P.}~\bibnamefont {Fierlinger}},
  \bibinfo {author} {\bibfnamefont {M.}~\bibnamefont {Ramsey-Musolf}}, \ and\
  \bibinfo {author} {\bibfnamefont {J.}~\bibnamefont {Singh}},\ }\href@noop {}
  {\bibfield  {journal} {\bibinfo  {journal} {arXiv preprint arXiv:1710.02504}\
  } (\bibinfo {year} {2017})}\BibitemShut {NoStop}%
\bibitem [{\citenamefont {Weinberg}(1978)}]{Weinberg:1977ma}%
  \BibitemOpen
  \bibfield  {author} {\bibinfo {author} {\bibfnamefont {S.}~\bibnamefont
  {Weinberg}},\ }\href {\doibase 10.1103/PhysRevLett.40.223} {\bibfield
  {journal} {\bibinfo  {journal} {Phys. Rev. Lett.}\ }\textbf {\bibinfo
  {volume} {40}},\ \bibinfo {pages} {223} (\bibinfo {year} {1978})}\BibitemShut
  {NoStop}%
\bibitem [{\citenamefont {Wilczek}(1978)}]{Wilczek:1977pj}%
  \BibitemOpen
  \bibfield  {author} {\bibinfo {author} {\bibfnamefont {F.}~\bibnamefont
  {Wilczek}},\ }\href {\doibase 10.1103/PhysRevLett.40.279} {\bibfield
  {journal} {\bibinfo  {journal} {Phys. Rev. Lett.}\ }\textbf {\bibinfo
  {volume} {40}},\ \bibinfo {pages} {279} (\bibinfo {year} {1978})}\BibitemShut
  {NoStop}%
\bibitem [{\citenamefont {Peccei}\ and\ \citenamefont
  {Quinn}(1977)}]{Peccei:1977hh}%
  \BibitemOpen
  \bibfield  {author} {\bibinfo {author} {\bibfnamefont {R.~D.}\ \bibnamefont
  {Peccei}}\ and\ \bibinfo {author} {\bibfnamefont {H.~R.}\ \bibnamefont
  {Quinn}},\ }\href {\doibase 10.1103/PhysRevLett.38.1440} {\bibfield
  {journal} {\bibinfo  {journal} {Phys. Rev. Lett.}\ }\textbf {\bibinfo
  {volume} {38}},\ \bibinfo {pages} {1440} (\bibinfo {year} {1977})},\ \bibinfo
  {note} {[,328(1977)]}\BibitemShut {NoStop}%
\bibitem [{\citenamefont {Graham}\ \emph {et~al.}(2015)\citenamefont {Graham},
  \citenamefont {Irastorza}, \citenamefont {Lamoreaux}, \citenamefont
  {Lindner},\ and\ \citenamefont {van Bibber}}]{Graham:2015ouw}%
  \BibitemOpen
  \bibfield  {author} {\bibinfo {author} {\bibfnamefont {P.~W.}\ \bibnamefont
  {Graham}}, \bibinfo {author} {\bibfnamefont {I.~G.}\ \bibnamefont
  {Irastorza}}, \bibinfo {author} {\bibfnamefont {S.~K.}\ \bibnamefont
  {Lamoreaux}}, \bibinfo {author} {\bibfnamefont {A.}~\bibnamefont {Lindner}},
  \ and\ \bibinfo {author} {\bibfnamefont {K.~A.}\ \bibnamefont {van Bibber}},\
  }\href {\doibase 10.1146/annurev-nucl-102014-022120} {\bibfield  {journal}
  {\bibinfo  {journal} {Ann. Rev. Nucl. Part. Sci.}\ }\textbf {\bibinfo
  {volume} {65}},\ \bibinfo {pages} {485} (\bibinfo {year} {2015})},\ \Eprint
  {http://arxiv.org/abs/1602.00039} {arXiv:1602.00039 [hep-ex]} \BibitemShut
  {NoStop}%
\bibitem [{\citenamefont {Mace}\ \emph {et~al.}(2016)\citenamefont {Mace},
  \citenamefont {Schlichting},\ and\ \citenamefont
  {Venugopalan}}]{Mace:2016svc}%
  \BibitemOpen
  \bibfield  {author} {\bibinfo {author} {\bibfnamefont {M.}~\bibnamefont
  {Mace}}, \bibinfo {author} {\bibfnamefont {S.}~\bibnamefont {Schlichting}}, \
  and\ \bibinfo {author} {\bibfnamefont {R.}~\bibnamefont {Venugopalan}},\
  }\href {\doibase 10.1103/PhysRevD.93.074036} {\bibfield  {journal} {\bibinfo
  {journal} {Phys. Rev.}\ }\textbf {\bibinfo {volume} {D93}},\ \bibinfo {pages}
  {074036} (\bibinfo {year} {2016})},\ \Eprint
  {http://arxiv.org/abs/1601.07342} {arXiv:1601.07342 [hep-ph]} \BibitemShut
  {NoStop}%
\bibitem [{\citenamefont {Cirac}\ and\ \citenamefont
  {Zoller}(2012)}]{cirac2012goals}%
  \BibitemOpen
  \bibfield  {author} {\bibinfo {author} {\bibfnamefont {J.~I.}\ \bibnamefont
  {Cirac}}\ and\ \bibinfo {author} {\bibfnamefont {P.}~\bibnamefont {Zoller}},\
  }\href@noop {} {\bibfield  {journal} {\bibinfo  {journal} {Nature Physics}\
  }\textbf {\bibinfo {volume} {8}},\ \bibinfo {pages} {264} (\bibinfo {year}
  {2012})}\BibitemShut {NoStop}%
\bibitem [{\citenamefont {Hauke}\ \emph {et~al.}(2012)\citenamefont {Hauke},
  \citenamefont {Cucchietti}, \citenamefont {Tagliacozzo}, \citenamefont
  {Deutsch},\ and\ \citenamefont {Lewenstein}}]{hauke2012can}%
  \BibitemOpen
  \bibfield  {author} {\bibinfo {author} {\bibfnamefont {P.}~\bibnamefont
  {Hauke}}, \bibinfo {author} {\bibfnamefont {F.~M.}\ \bibnamefont
  {Cucchietti}}, \bibinfo {author} {\bibfnamefont {L.}~\bibnamefont
  {Tagliacozzo}}, \bibinfo {author} {\bibfnamefont {I.}~\bibnamefont
  {Deutsch}}, \ and\ \bibinfo {author} {\bibfnamefont {M.}~\bibnamefont
  {Lewenstein}},\ }\href@noop {} {\bibfield  {journal} {\bibinfo  {journal}
  {Reports on Progress in Physics}\ }\textbf {\bibinfo {volume} {75}},\
  \bibinfo {pages} {082401} (\bibinfo {year} {2012})}\BibitemShut {NoStop}%
\bibitem [{\citenamefont {Carlson}\ \emph {et~al.}()\citenamefont {Carlson},
  \citenamefont {Dean}, \citenamefont {M.}, \citenamefont {Kaplan},
  \citenamefont {Preskill}, \citenamefont {Roche}, \citenamefont {M.},\ and\
  \citenamefont {Troyer}}]{nuclearReview}%
  \BibitemOpen
  \bibfield  {author} {\bibinfo {author} {\bibfnamefont {J.}~\bibnamefont
  {Carlson}}, \bibinfo {author} {\bibfnamefont {D.}~\bibnamefont {Dean}},
  \bibinfo {author} {\bibfnamefont {H.-J.}\ \bibnamefont {M.}}, \bibinfo
  {author} {\bibfnamefont {D.}~\bibnamefont {Kaplan}}, \bibinfo {author}
  {\bibfnamefont {J.}~\bibnamefont {Preskill}}, \bibinfo {author}
  {\bibfnamefont {K.}~\bibnamefont {Roche}}, \bibinfo {author} {\bibfnamefont
  {S.}~\bibnamefont {M.}}, \ and\ \bibinfo {author} {\bibfnamefont
  {M.}~\bibnamefont {Troyer}},\ }\href@noop {} {\bibinfo  {journal} {Institute
  For Nuclear Theory Report 18-008}\ }\BibitemShut {NoStop}%
\bibitem [{\citenamefont {Martinez}\ \emph {et~al.}(2016)\citenamefont
  {Martinez}, \citenamefont {Muschik}, \citenamefont {Schindler}, \citenamefont
  {Nigg}, \citenamefont {Erhard}, \citenamefont {Heyl}, \citenamefont {Hauke},
  \citenamefont {Dalmonte}, \citenamefont {Monz}, \citenamefont {Zoller} \emph
  {et~al.}}]{martinez2016real}%
  \BibitemOpen
\bibfield  {journal} {  }\bibfield  {author} {\bibinfo {author} {\bibfnamefont
  {E.~A.}\ \bibnamefont {Martinez}}, \bibinfo {author} {\bibfnamefont {C.~A.}\
  \bibnamefont {Muschik}}, \bibinfo {author} {\bibfnamefont {P.}~\bibnamefont
  {Schindler}}, \bibinfo {author} {\bibfnamefont {D.}~\bibnamefont {Nigg}},
  \bibinfo {author} {\bibfnamefont {A.}~\bibnamefont {Erhard}}, \bibinfo
  {author} {\bibfnamefont {M.}~\bibnamefont {Heyl}}, \bibinfo {author}
  {\bibfnamefont {P.}~\bibnamefont {Hauke}}, \bibinfo {author} {\bibfnamefont
  {M.}~\bibnamefont {Dalmonte}}, \bibinfo {author} {\bibfnamefont
  {T.}~\bibnamefont {Monz}}, \bibinfo {author} {\bibfnamefont {P.}~\bibnamefont
  {Zoller}},  \emph {et~al.},\ }\href@noop {} {\bibfield  {journal} {\bibinfo
  {journal} {Nature}\ }\textbf {\bibinfo {volume} {534}},\ \bibinfo {pages}
  {516} (\bibinfo {year} {2016})}\BibitemShut {NoStop}%
\bibitem [{\citenamefont {Klco}\ \emph {et~al.}(2018)\citenamefont {Klco},
  \citenamefont {Dumitrescu}, \citenamefont {McCaskey}, \citenamefont {Morris},
  \citenamefont {Pooser}, \citenamefont {Sanz}, \citenamefont {Solano},
  \citenamefont {Lougovski},\ and\ \citenamefont {Savage}}]{klco2018quantum}%
  \BibitemOpen
  \bibfield  {author} {\bibinfo {author} {\bibfnamefont {N.}~\bibnamefont
  {Klco}}, \bibinfo {author} {\bibfnamefont {E.}~\bibnamefont {Dumitrescu}},
  \bibinfo {author} {\bibfnamefont {A.}~\bibnamefont {McCaskey}}, \bibinfo
  {author} {\bibfnamefont {T.}~\bibnamefont {Morris}}, \bibinfo {author}
  {\bibfnamefont {R.}~\bibnamefont {Pooser}}, \bibinfo {author} {\bibfnamefont
  {M.}~\bibnamefont {Sanz}}, \bibinfo {author} {\bibfnamefont {E.}~\bibnamefont
  {Solano}}, \bibinfo {author} {\bibfnamefont {P.}~\bibnamefont {Lougovski}}, \
  and\ \bibinfo {author} {\bibfnamefont {M.}~\bibnamefont {Savage}},\
  }\href@noop {} {\bibfield  {journal} {\bibinfo  {journal} {arXiv preprint
  arXiv:1803.03326}\ } (\bibinfo {year} {2018})}\BibitemShut {NoStop}%
\bibitem [{\citenamefont {Coleman}(1976)}]{Coleman:1976uz}%
  \BibitemOpen
  \bibfield  {author} {\bibinfo {author} {\bibfnamefont {S.~R.}\ \bibnamefont
  {Coleman}},\ }\href {\doibase 10.1016/0003-4916(76)90280-3} {\bibfield
  {journal} {\bibinfo  {journal} {Annals Phys.}\ }\textbf {\bibinfo {volume}
  {101}},\ \bibinfo {pages} {239} (\bibinfo {year} {1976})}\BibitemShut
  {NoStop}%
\bibitem [{\citenamefont {Coleman}\ \emph {et~al.}(1975)\citenamefont
  {Coleman}, \citenamefont {Jackiw},\ and\ \citenamefont
  {Susskind}}]{coleman1975charge}%
  \BibitemOpen
  \bibfield  {author} {\bibinfo {author} {\bibfnamefont {S.}~\bibnamefont
  {Coleman}}, \bibinfo {author} {\bibfnamefont {R.}~\bibnamefont {Jackiw}}, \
  and\ \bibinfo {author} {\bibfnamefont {L.}~\bibnamefont {Susskind}},\
  }\href@noop {} {\bibfield  {journal} {\bibinfo  {journal} {Annals of
  Physics}\ }\textbf {\bibinfo {volume} {93}},\ \bibinfo {pages} {267}
  (\bibinfo {year} {1975})}\BibitemShut {NoStop}%
\bibitem [{\citenamefont {Petreczky}(2012)}]{Petreczky:2012rq}%
  \BibitemOpen
  \bibfield  {author} {\bibinfo {author} {\bibfnamefont {P.}~\bibnamefont
  {Petreczky}},\ }\href {\doibase 10.1088/0954-3899/39/9/093002} {\bibfield
  {journal} {\bibinfo  {journal} {J. Phys.}\ }\textbf {\bibinfo {volume}
  {G39}},\ \bibinfo {pages} {093002} (\bibinfo {year} {2012})},\ \Eprint
  {http://arxiv.org/abs/1203.5320} {arXiv:1203.5320 [hep-lat]} \BibitemShut
  {NoStop}%
\bibitem [{\citenamefont {Heyl}\ \emph {et~al.}(2013)\citenamefont {Heyl},
  \citenamefont {Polkovnikov},\ and\ \citenamefont
  {Kehrein}}]{heyl2013dynamical}%
  \BibitemOpen
  \bibfield  {author} {\bibinfo {author} {\bibfnamefont {M.}~\bibnamefont
  {Heyl}}, \bibinfo {author} {\bibfnamefont {A.}~\bibnamefont {Polkovnikov}}, \
  and\ \bibinfo {author} {\bibfnamefont {S.}~\bibnamefont {Kehrein}},\
  }\href@noop {} {\bibfield  {journal} {\bibinfo  {journal} {Physical review
  letters}\ }\textbf {\bibinfo {volume} {110}},\ \bibinfo {pages} {135704}
  (\bibinfo {year} {2013})}\BibitemShut {NoStop}%
\bibitem [{\citenamefont {Fl{\"a}schner}\ \emph {et~al.}(2016)\citenamefont
  {Fl{\"a}schner}, \citenamefont {Vogel}, \citenamefont {Tarnowski},
  \citenamefont {Rem}, \citenamefont {L{\"u}hmann}, \citenamefont {Heyl},
  \citenamefont {Budich}, \citenamefont {Mathey}, \citenamefont {Sengstock},\
  and\ \citenamefont {Weitenberg}}]{flaschner2016observation}%
  \BibitemOpen
  \bibfield  {author} {\bibinfo {author} {\bibfnamefont {N.}~\bibnamefont
  {Fl{\"a}schner}}, \bibinfo {author} {\bibfnamefont {D.}~\bibnamefont
  {Vogel}}, \bibinfo {author} {\bibfnamefont {M.}~\bibnamefont {Tarnowski}},
  \bibinfo {author} {\bibfnamefont {B.~S.}\ \bibnamefont {Rem}}, \bibinfo
  {author} {\bibfnamefont {D.-S.}\ \bibnamefont {L{\"u}hmann}}, \bibinfo
  {author} {\bibfnamefont {M.}~\bibnamefont {Heyl}}, \bibinfo {author}
  {\bibfnamefont {J.~C.}\ \bibnamefont {Budich}}, \bibinfo {author}
  {\bibfnamefont {L.}~\bibnamefont {Mathey}}, \bibinfo {author} {\bibfnamefont
  {K.}~\bibnamefont {Sengstock}}, \ and\ \bibinfo {author} {\bibfnamefont
  {C.}~\bibnamefont {Weitenberg}},\ }\href@noop {} {\bibfield  {journal}
  {\bibinfo  {journal} {arXiv preprint arXiv:1608.05616}\ } (\bibinfo {year}
  {2016})}\BibitemShut {NoStop}%
\bibitem [{\citenamefont {Jurcevic}\ \emph {et~al.}(2017)\citenamefont
  {Jurcevic}, \citenamefont {Shen}, \citenamefont {Hauke}, \citenamefont
  {Maier}, \citenamefont {Brydges}, \citenamefont {Hempel}, \citenamefont
  {Lanyon}, \citenamefont {Heyl}, \citenamefont {Blatt},\ and\ \citenamefont
  {Roos}}]{jurcevic2017direct}%
  \BibitemOpen
  \bibfield  {author} {\bibinfo {author} {\bibfnamefont {P.}~\bibnamefont
  {Jurcevic}}, \bibinfo {author} {\bibfnamefont {H.}~\bibnamefont {Shen}},
  \bibinfo {author} {\bibfnamefont {P.}~\bibnamefont {Hauke}}, \bibinfo
  {author} {\bibfnamefont {C.}~\bibnamefont {Maier}}, \bibinfo {author}
  {\bibfnamefont {T.}~\bibnamefont {Brydges}}, \bibinfo {author} {\bibfnamefont
  {C.}~\bibnamefont {Hempel}}, \bibinfo {author} {\bibfnamefont
  {B.}~\bibnamefont {Lanyon}}, \bibinfo {author} {\bibfnamefont
  {M.}~\bibnamefont {Heyl}}, \bibinfo {author} {\bibfnamefont {R.}~\bibnamefont
  {Blatt}}, \ and\ \bibinfo {author} {\bibfnamefont {C.}~\bibnamefont {Roos}},\
  }\href@noop {} {\bibfield  {journal} {\bibinfo  {journal} {Physical review
  letters}\ }\textbf {\bibinfo {volume} {119}},\ \bibinfo {pages} {080501}
  (\bibinfo {year} {2017})}\BibitemShut {NoStop}%
\bibitem [{\citenamefont {Heyl}(2018)}]{heyl2018dynamical}%
  \BibitemOpen
  \bibfield  {author} {\bibinfo {author} {\bibfnamefont {M.}~\bibnamefont
  {Heyl}},\ }\href@noop {} {\bibfield  {journal} {\bibinfo  {journal} {Reports
  on Progress in Physics}\ }\textbf {\bibinfo {volume} {81}},\ \bibinfo {pages}
  {054001} (\bibinfo {year} {2018})}\BibitemShut {NoStop}%
\bibitem [{\citenamefont {Budich}\ and\ \citenamefont
  {Heyl}(2016)}]{budich2016dynamical}%
  \BibitemOpen
  \bibfield  {author} {\bibinfo {author} {\bibfnamefont {J.~C.}\ \bibnamefont
  {Budich}}\ and\ \bibinfo {author} {\bibfnamefont {M.}~\bibnamefont {Heyl}},\
  }\href@noop {} {\bibfield  {journal} {\bibinfo  {journal} {Physical Review
  B}\ }\textbf {\bibinfo {volume} {93}},\ \bibinfo {pages} {085416} (\bibinfo
  {year} {2016})}\BibitemShut {NoStop}%
\bibitem [{\citenamefont {Tian}\ \emph {et~al.}(2018)\citenamefont {Tian},
  \citenamefont {Ke}, \citenamefont {Zhang}, \citenamefont {Lin}, \citenamefont
  {Shi}, \citenamefont {Huang}, \citenamefont {Lee},\ and\ \citenamefont
  {Du}}]{tian2018direct}%
  \BibitemOpen
  \bibfield  {author} {\bibinfo {author} {\bibfnamefont {T.}~\bibnamefont
  {Tian}}, \bibinfo {author} {\bibfnamefont {Y.}~\bibnamefont {Ke}}, \bibinfo
  {author} {\bibfnamefont {L.}~\bibnamefont {Zhang}}, \bibinfo {author}
  {\bibfnamefont {S.}~\bibnamefont {Lin}}, \bibinfo {author} {\bibfnamefont
  {Z.}~\bibnamefont {Shi}}, \bibinfo {author} {\bibfnamefont {P.}~\bibnamefont
  {Huang}}, \bibinfo {author} {\bibfnamefont {C.}~\bibnamefont {Lee}}, \ and\
  \bibinfo {author} {\bibfnamefont {J.}~\bibnamefont {Du}},\ }\href@noop {}
  {\bibfield  {journal} {\bibinfo  {journal} {arXiv preprint arXiv:1807.04483}\
  } (\bibinfo {year} {2018})}\BibitemShut {NoStop}%
\bibitem [{\citenamefont {Xu}\ \emph {et~al.}(2018)\citenamefont {Xu},
  \citenamefont {Wang}, \citenamefont {Heyl}, \citenamefont {Budich},
  \citenamefont {Pan}, \citenamefont {Chen}, \citenamefont {Jan}, \citenamefont
  {Sun}, \citenamefont {Xu}, \citenamefont {Han} \emph
  {et~al.}}]{xu2018measuring}%
  \BibitemOpen
  \bibfield  {author} {\bibinfo {author} {\bibfnamefont {X.-Y.}\ \bibnamefont
  {Xu}}, \bibinfo {author} {\bibfnamefont {Q.-Q.}\ \bibnamefont {Wang}},
  \bibinfo {author} {\bibfnamefont {M.}~\bibnamefont {Heyl}}, \bibinfo {author}
  {\bibfnamefont {J.~C.}\ \bibnamefont {Budich}}, \bibinfo {author}
  {\bibfnamefont {W.-W.}\ \bibnamefont {Pan}}, \bibinfo {author} {\bibfnamefont
  {Z.}~\bibnamefont {Chen}}, \bibinfo {author} {\bibfnamefont {M.}~\bibnamefont
  {Jan}}, \bibinfo {author} {\bibfnamefont {K.}~\bibnamefont {Sun}}, \bibinfo
  {author} {\bibfnamefont {J.-S.}\ \bibnamefont {Xu}}, \bibinfo {author}
  {\bibfnamefont {Y.-J.}\ \bibnamefont {Han}},  \emph {et~al.},\ }\href@noop {}
  {\bibfield  {journal} {\bibinfo  {journal} {arXiv preprint arXiv:1808.03930}\
  } (\bibinfo {year} {2018})}\BibitemShut {NoStop}%
\bibitem [{\citenamefont {Zache}\ \emph {et~al.}(2018)\citenamefont {Zache},
  \citenamefont {Hebenstreit}, \citenamefont {Jendrzejewski}, \citenamefont
  {Oberthaler}, \citenamefont {Berges},\ and\ \citenamefont
  {Hauke}}]{zache2018quantum}%
  \BibitemOpen
  \bibfield  {author} {\bibinfo {author} {\bibfnamefont {T.~V.}\ \bibnamefont
  {Zache}}, \bibinfo {author} {\bibfnamefont {F.}~\bibnamefont {Hebenstreit}},
  \bibinfo {author} {\bibfnamefont {F.}~\bibnamefont {Jendrzejewski}}, \bibinfo
  {author} {\bibfnamefont {M.}~\bibnamefont {Oberthaler}}, \bibinfo {author}
  {\bibfnamefont {J.}~\bibnamefont {Berges}}, \ and\ \bibinfo {author}
  {\bibfnamefont {P.}~\bibnamefont {Hauke}},\ }\href@noop {} {\bibfield
  {journal} {\bibinfo  {journal} {Quantum Science and Technology}\ } (\bibinfo
  {year} {2018})}\BibitemShut {NoStop}%
\bibitem [{\citenamefont {Gelis}\ and\ \citenamefont
  {Tanji}(2016)}]{Gelis:2015kya}%
  \BibitemOpen
  \bibfield  {author} {\bibinfo {author} {\bibfnamefont {F.}~\bibnamefont
  {Gelis}}\ and\ \bibinfo {author} {\bibfnamefont {N.}~\bibnamefont {Tanji}},\
  }\href {\doibase 10.1016/j.ppnp.2015.11.001} {\bibfield  {journal} {\bibinfo
  {journal} {Prog. Part. Nucl. Phys.}\ }\textbf {\bibinfo {volume} {87}},\
  \bibinfo {pages} {1} (\bibinfo {year} {2016})},\ \Eprint
  {http://arxiv.org/abs/1510.05451} {arXiv:1510.05451 [hep-ph]} \BibitemShut
  {NoStop}%
\bibitem [{\citenamefont {Gorin}\ \emph {et~al.}(2006)\citenamefont {Gorin},
  \citenamefont {Prosen}, \citenamefont {Seligman},\ and\ \citenamefont
  {{\v{Z}}nidari{\v{c}}}}]{gorin2006dynamics}%
  \BibitemOpen
  \bibfield  {author} {\bibinfo {author} {\bibfnamefont {T.}~\bibnamefont
  {Gorin}}, \bibinfo {author} {\bibfnamefont {T.}~\bibnamefont {Prosen}},
  \bibinfo {author} {\bibfnamefont {T.~H.}\ \bibnamefont {Seligman}}, \ and\
  \bibinfo {author} {\bibfnamefont {M.}~\bibnamefont {{\v{Z}}nidari{\v{c}}}},\
  }\href@noop {} {\bibfield  {journal} {\bibinfo  {journal} {Physics Reports}\
  }\textbf {\bibinfo {volume} {435}},\ \bibinfo {pages} {33} (\bibinfo {year}
  {2006})}\BibitemShut {NoStop}%
\bibitem [{\citenamefont {Banks}\ \emph {et~al.}(1976)\citenamefont {Banks},
  \citenamefont {Susskind},\ and\ \citenamefont {Kogut}}]{Banks:1975gq}%
  \BibitemOpen
  \bibfield  {author} {\bibinfo {author} {\bibfnamefont {T.}~\bibnamefont
  {Banks}}, \bibinfo {author} {\bibfnamefont {L.}~\bibnamefont {Susskind}}, \
  and\ \bibinfo {author} {\bibfnamefont {J.~B.}\ \bibnamefont {Kogut}},\ }\href
  {\doibase 10.1103/PhysRevD.13.1043} {\bibfield  {journal} {\bibinfo
  {journal} {Phys. Rev.}\ }\textbf {\bibinfo {volume} {D13}},\ \bibinfo {pages}
  {1043} (\bibinfo {year} {1976})}\BibitemShut {NoStop}%
\bibitem [{\citenamefont {Hamer}\ \emph {et~al.}(1997)\citenamefont {Hamer},
  \citenamefont {Weihong},\ and\ \citenamefont {Oitmaa}}]{hamer1997series}%
  \BibitemOpen
  \bibfield  {author} {\bibinfo {author} {\bibfnamefont {C.}~\bibnamefont
  {Hamer}}, \bibinfo {author} {\bibfnamefont {Z.}~\bibnamefont {Weihong}}, \
  and\ \bibinfo {author} {\bibfnamefont {J.}~\bibnamefont {Oitmaa}},\
  }\href@noop {} {\bibfield  {journal} {\bibinfo  {journal} {Physical Review
  D}\ }\textbf {\bibinfo {volume} {56}},\ \bibinfo {pages} {55} (\bibinfo
  {year} {1997})}\BibitemShut {NoStop}%
\bibitem [{lon()}]{longpaper}%
  \BibitemOpen
  \href@noop {} {}\bibinfo {note} {T. V. Zache et al., in
  preparation}\BibitemShut {NoStop}%
\bibitem [{\citenamefont {Fukui}\ \emph {et~al.}(2005)\citenamefont {Fukui},
  \citenamefont {Hatsugai},\ and\ \citenamefont {Suzuki}}]{fukui2005chern}%
  \BibitemOpen
  \bibfield  {author} {\bibinfo {author} {\bibfnamefont {T.}~\bibnamefont
  {Fukui}}, \bibinfo {author} {\bibfnamefont {Y.}~\bibnamefont {Hatsugai}}, \
  and\ \bibinfo {author} {\bibfnamefont {H.}~\bibnamefont {Suzuki}},\
  }\href@noop {} {\bibfield  {journal} {\bibinfo  {journal} {Journal of the
  Physical Society of Japan}\ }\textbf {\bibinfo {volume} {74}},\ \bibinfo
  {pages} {1674} (\bibinfo {year} {2005})}\BibitemShut {NoStop}%
\bibitem [{\citenamefont {Abdalla}\ \emph {et~al.}(1991)\citenamefont
  {Abdalla}, \citenamefont {Abdalla},\ and\ \citenamefont
  {Rothe}}]{abdalla1991non}%
  \BibitemOpen
  \bibfield  {author} {\bibinfo {author} {\bibfnamefont {E.}~\bibnamefont
  {Abdalla}}, \bibinfo {author} {\bibfnamefont {M.~C.~B.}\ \bibnamefont
  {Abdalla}}, \ and\ \bibinfo {author} {\bibfnamefont {K.~D.}\ \bibnamefont
  {Rothe}},\ }\href@noop {} {\emph {\bibinfo {title} {Non-perturbative methods
  in 2 dimensional quantum field theory}}}\ (\bibinfo  {publisher} {World
  Scientific},\ \bibinfo {year} {1991})\BibitemShut {NoStop}%
\bibitem [{\citenamefont {Monz}\ \emph {et~al.}(2016)\citenamefont {Monz},
  \citenamefont {Nigg}, \citenamefont {Martinez}, \citenamefont {Brandl},
  \citenamefont {Schindler}, \citenamefont {Rines}, \citenamefont {Wang},
  \citenamefont {Chuang},\ and\ \citenamefont {Blatt}}]{monz2016realization}%
  \BibitemOpen
  \bibfield  {author} {\bibinfo {author} {\bibfnamefont {T.}~\bibnamefont
  {Monz}}, \bibinfo {author} {\bibfnamefont {D.}~\bibnamefont {Nigg}}, \bibinfo
  {author} {\bibfnamefont {E.~A.}\ \bibnamefont {Martinez}}, \bibinfo {author}
  {\bibfnamefont {M.~F.}\ \bibnamefont {Brandl}}, \bibinfo {author}
  {\bibfnamefont {P.}~\bibnamefont {Schindler}}, \bibinfo {author}
  {\bibfnamefont {R.}~\bibnamefont {Rines}}, \bibinfo {author} {\bibfnamefont
  {S.~X.}\ \bibnamefont {Wang}}, \bibinfo {author} {\bibfnamefont {I.~L.}\
  \bibnamefont {Chuang}}, \ and\ \bibinfo {author} {\bibfnamefont
  {R.}~\bibnamefont {Blatt}},\ }\href@noop {} {\bibfield  {journal} {\bibinfo
  {journal} {Science}\ }\textbf {\bibinfo {volume} {351}},\ \bibinfo {pages}
  {1068} (\bibinfo {year} {2016})}\BibitemShut {NoStop}%
\bibitem [{\citenamefont {Barends}\ \emph {et~al.}(2016)\citenamefont
  {Barends}, \citenamefont {Shabani}, \citenamefont {Lamata}, \citenamefont
  {Kelly}, \citenamefont {Mezzacapo}, \citenamefont {Las~Heras}, \citenamefont
  {Babbush}, \citenamefont {Fowler}, \citenamefont {Campbell}, \citenamefont
  {Chen} \emph {et~al.}}]{barends2016digitized}%
  \BibitemOpen
  \bibfield  {author} {\bibinfo {author} {\bibfnamefont {R.}~\bibnamefont
  {Barends}}, \bibinfo {author} {\bibfnamefont {A.}~\bibnamefont {Shabani}},
  \bibinfo {author} {\bibfnamefont {L.}~\bibnamefont {Lamata}}, \bibinfo
  {author} {\bibfnamefont {J.}~\bibnamefont {Kelly}}, \bibinfo {author}
  {\bibfnamefont {A.}~\bibnamefont {Mezzacapo}}, \bibinfo {author}
  {\bibfnamefont {U.}~\bibnamefont {Las~Heras}}, \bibinfo {author}
  {\bibfnamefont {R.}~\bibnamefont {Babbush}}, \bibinfo {author} {\bibfnamefont
  {A.~G.}\ \bibnamefont {Fowler}}, \bibinfo {author} {\bibfnamefont
  {B.}~\bibnamefont {Campbell}}, \bibinfo {author} {\bibfnamefont
  {Y.}~\bibnamefont {Chen}},  \emph {et~al.},\ }\href@noop {} {\bibfield
  {journal} {\bibinfo  {journal} {Nature}\ }\textbf {\bibinfo {volume} {534}},\
  \bibinfo {pages} {222} (\bibinfo {year} {2016})}\BibitemShut {NoStop}%
\bibitem [{\citenamefont {Kandala}\ \emph {et~al.}(2017)\citenamefont
  {Kandala}, \citenamefont {Mezzacapo}, \citenamefont {Temme}, \citenamefont
  {Takita}, \citenamefont {Brink}, \citenamefont {Chow},\ and\ \citenamefont
  {Gambetta}}]{kandala2017hardware}%
  \BibitemOpen
  \bibfield  {author} {\bibinfo {author} {\bibfnamefont {A.}~\bibnamefont
  {Kandala}}, \bibinfo {author} {\bibfnamefont {A.}~\bibnamefont {Mezzacapo}},
  \bibinfo {author} {\bibfnamefont {K.}~\bibnamefont {Temme}}, \bibinfo
  {author} {\bibfnamefont {M.}~\bibnamefont {Takita}}, \bibinfo {author}
  {\bibfnamefont {M.}~\bibnamefont {Brink}}, \bibinfo {author} {\bibfnamefont
  {J.~M.}\ \bibnamefont {Chow}}, \ and\ \bibinfo {author} {\bibfnamefont
  {J.~M.}\ \bibnamefont {Gambetta}},\ }\href@noop {} {\bibfield  {journal}
  {\bibinfo  {journal} {Nature}\ }\textbf {\bibinfo {volume} {549}},\ \bibinfo
  {pages} {242} (\bibinfo {year} {2017})}\BibitemShut {NoStop}%
\bibitem [{\citenamefont {Landsman}\ \emph {et~al.}(2018)\citenamefont
  {Landsman}, \citenamefont {Figgatt}, \citenamefont {Schuster}, \citenamefont
  {Linke}, \citenamefont {Yoshida}, \citenamefont {Yao},\ and\ \citenamefont
  {Monroe}}]{landsman2018verified}%
  \BibitemOpen
  \bibfield  {author} {\bibinfo {author} {\bibfnamefont {K.~A.}\ \bibnamefont
  {Landsman}}, \bibinfo {author} {\bibfnamefont {C.}~\bibnamefont {Figgatt}},
  \bibinfo {author} {\bibfnamefont {T.}~\bibnamefont {Schuster}}, \bibinfo
  {author} {\bibfnamefont {N.~M.}\ \bibnamefont {Linke}}, \bibinfo {author}
  {\bibfnamefont {B.}~\bibnamefont {Yoshida}}, \bibinfo {author} {\bibfnamefont
  {N.~Y.}\ \bibnamefont {Yao}}, \ and\ \bibinfo {author} {\bibfnamefont
  {C.}~\bibnamefont {Monroe}},\ }\href@noop {} {\bibfield  {journal} {\bibinfo
  {journal} {arXiv preprint arXiv:1806.02807}\ } (\bibinfo {year}
  {2018})}\BibitemShut {NoStop}%
\bibitem [{\citenamefont {Wiese}(2013)}]{wiese2013ultracold}%
  \BibitemOpen
  \bibfield  {author} {\bibinfo {author} {\bibfnamefont {U.-J.}\ \bibnamefont
  {Wiese}},\ }\href@noop {} {\bibfield  {journal} {\bibinfo  {journal} {Annalen
  der Physik}\ }\textbf {\bibinfo {volume} {525}},\ \bibinfo {pages} {777}
  (\bibinfo {year} {2013})}\BibitemShut {NoStop}%
\bibitem [{\citenamefont {Zohar}\ \emph {et~al.}(2015)\citenamefont {Zohar},
  \citenamefont {Cirac},\ and\ \citenamefont {Reznik}}]{zohar2015quantum}%
  \BibitemOpen
  \bibfield  {author} {\bibinfo {author} {\bibfnamefont {E.}~\bibnamefont
  {Zohar}}, \bibinfo {author} {\bibfnamefont {J.~I.}\ \bibnamefont {Cirac}}, \
  and\ \bibinfo {author} {\bibfnamefont {B.}~\bibnamefont {Reznik}},\
  }\href@noop {} {\bibfield  {journal} {\bibinfo  {journal} {Reports on
  Progress in Physics}\ }\textbf {\bibinfo {volume} {79}},\ \bibinfo {pages}
  {014401} (\bibinfo {year} {2015})}\BibitemShut {NoStop}%
\bibitem [{\citenamefont {Dalmonte}\ and\ \citenamefont
  {Montangero}(2016)}]{dalmonte2016lattice}%
  \BibitemOpen
  \bibfield  {author} {\bibinfo {author} {\bibfnamefont {M.}~\bibnamefont
  {Dalmonte}}\ and\ \bibinfo {author} {\bibfnamefont {S.}~\bibnamefont
  {Montangero}},\ }\href@noop {} {\bibfield  {journal} {\bibinfo  {journal}
  {Contemporary Physics}\ }\textbf {\bibinfo {volume} {57}},\ \bibinfo {pages}
  {388} (\bibinfo {year} {2016})}\BibitemShut {NoStop}%
\bibitem [{\citenamefont {Magnifico}\ \emph {et~al.}(2018)\citenamefont
  {Magnifico}, \citenamefont {Vodola}, \citenamefont {Ercolessi}, \citenamefont
  {Kumar}, \citenamefont {Muller},\ and\ \citenamefont
  {Bermudez}}]{Magnifico:2018wek}%
  \BibitemOpen
  \bibfield  {author} {\bibinfo {author} {\bibfnamefont {G.}~\bibnamefont
  {Magnifico}}, \bibinfo {author} {\bibfnamefont {D.}~\bibnamefont {Vodola}},
  \bibinfo {author} {\bibfnamefont {E.}~\bibnamefont {Ercolessi}}, \bibinfo
  {author} {\bibfnamefont {S.~P.}\ \bibnamefont {Kumar}}, \bibinfo {author}
  {\bibfnamefont {M.}~\bibnamefont {Muller}}, \ and\ \bibinfo {author}
  {\bibfnamefont {A.}~\bibnamefont {Bermudez}},\ }\href@noop {} {\  (\bibinfo
  {year} {2018})},\ \Eprint {http://arxiv.org/abs/1804.10568} {arXiv:1804.10568
  [cond-mat.quant-gas]} \BibitemShut {NoStop}%
\bibitem [{\citenamefont {Knap}\ \emph {et~al.}(2013)\citenamefont {Knap},
  \citenamefont {Kantian}, \citenamefont {Giamarchi}, \citenamefont {Bloch},
  \citenamefont {Lukin},\ and\ \citenamefont {Demler}}]{knap2013probing}%
  \BibitemOpen
  \bibfield  {author} {\bibinfo {author} {\bibfnamefont {M.}~\bibnamefont
  {Knap}}, \bibinfo {author} {\bibfnamefont {A.}~\bibnamefont {Kantian}},
  \bibinfo {author} {\bibfnamefont {T.}~\bibnamefont {Giamarchi}}, \bibinfo
  {author} {\bibfnamefont {I.}~\bibnamefont {Bloch}}, \bibinfo {author}
  {\bibfnamefont {M.~D.}\ \bibnamefont {Lukin}}, \ and\ \bibinfo {author}
  {\bibfnamefont {E.}~\bibnamefont {Demler}},\ }\href@noop {} {\bibfield
  {journal} {\bibinfo  {journal} {Physical review letters}\ }\textbf {\bibinfo
  {volume} {111}},\ \bibinfo {pages} {147205} (\bibinfo {year}
  {2013})}\BibitemShut {NoStop}%
\bibitem [{\citenamefont {Uhrich}\ \emph {et~al.}(2018)\citenamefont {Uhrich},
  \citenamefont {Gross},\ and\ \citenamefont {Kastner}}]{uhrich2018probing}%
  \BibitemOpen
  \bibfield  {author} {\bibinfo {author} {\bibfnamefont {P.}~\bibnamefont
  {Uhrich}}, \bibinfo {author} {\bibfnamefont {C.}~\bibnamefont {Gross}}, \
  and\ \bibinfo {author} {\bibfnamefont {M.}~\bibnamefont {Kastner}},\
  }\href@noop {} {\bibfield  {journal} {\bibinfo  {journal} {arXiv preprint
  arXiv:1806.01758}\ } (\bibinfo {year} {2018})}\BibitemShut {NoStop}%
\bibitem [{\citenamefont {Gurarie}(2011)}]{gurarie2011single}%
  \BibitemOpen
  \bibfield  {author} {\bibinfo {author} {\bibfnamefont {V.}~\bibnamefont
  {Gurarie}},\ }\href@noop {} {\bibfield  {journal} {\bibinfo  {journal}
  {Physical Review B}\ }\textbf {\bibinfo {volume} {83}},\ \bibinfo {pages}
  {085426} (\bibinfo {year} {2011})}\BibitemShut {NoStop}%
\bibitem [{\citenamefont {Rachel}(2018)}]{rachel2018interacting}%
  \BibitemOpen
  \bibfield  {author} {\bibinfo {author} {\bibfnamefont {S.}~\bibnamefont
  {Rachel}},\ }\href@noop {} {\bibfield  {journal} {\bibinfo  {journal} {arXiv
  preprint arXiv:1804.10656}\ } (\bibinfo {year} {2018})}\BibitemShut {NoStop}%
\bibitem [{\citenamefont {Kharzeev}\ \emph {et~al.}(2008)\citenamefont
  {Kharzeev}, \citenamefont {McLerran},\ and\ \citenamefont
  {Warringa}}]{Kharzeev:2007jp}%
  \BibitemOpen
  \bibfield  {author} {\bibinfo {author} {\bibfnamefont {D.~E.}\ \bibnamefont
  {Kharzeev}}, \bibinfo {author} {\bibfnamefont {L.~D.}\ \bibnamefont
  {McLerran}}, \ and\ \bibinfo {author} {\bibfnamefont {H.~J.}\ \bibnamefont
  {Warringa}},\ }\href {\doibase 10.1016/j.nuclphysa.2008.02.298} {\bibfield
  {journal} {\bibinfo  {journal} {Nucl. Phys.}\ }\textbf {\bibinfo {volume}
  {A803}},\ \bibinfo {pages} {227} (\bibinfo {year} {2008})},\ \Eprint
  {http://arxiv.org/abs/0711.0950} {arXiv:0711.0950 [hep-ph]} \BibitemShut
  {NoStop}%
\bibitem [{\citenamefont {Fukushima}\ \emph {et~al.}(2008)\citenamefont
  {Fukushima}, \citenamefont {Kharzeev},\ and\ \citenamefont
  {Warringa}}]{Fukushima:2008xe}%
  \BibitemOpen
  \bibfield  {author} {\bibinfo {author} {\bibfnamefont {K.}~\bibnamefont
  {Fukushima}}, \bibinfo {author} {\bibfnamefont {D.~E.}\ \bibnamefont
  {Kharzeev}}, \ and\ \bibinfo {author} {\bibfnamefont {H.~J.}\ \bibnamefont
  {Warringa}},\ }\href {\doibase 10.1103/PhysRevD.78.074033} {\bibfield
  {journal} {\bibinfo  {journal} {Phys. Rev.}\ }\textbf {\bibinfo {volume}
  {D78}},\ \bibinfo {pages} {074033} (\bibinfo {year} {2008})},\ \Eprint
  {http://arxiv.org/abs/0808.3382} {arXiv:0808.3382 [hep-ph]} \BibitemShut
  {NoStop}%
\bibitem [{\citenamefont {Kharzeev}\ \emph {et~al.}(2016)\citenamefont
  {Kharzeev}, \citenamefont {Liao}, \citenamefont {Voloshin},\ and\
  \citenamefont {Wang}}]{Kharzeev:2015znc}%
  \BibitemOpen
  \bibfield  {author} {\bibinfo {author} {\bibfnamefont {D.~E.}\ \bibnamefont
  {Kharzeev}}, \bibinfo {author} {\bibfnamefont {J.}~\bibnamefont {Liao}},
  \bibinfo {author} {\bibfnamefont {S.~A.}\ \bibnamefont {Voloshin}}, \ and\
  \bibinfo {author} {\bibfnamefont {G.}~\bibnamefont {Wang}},\ }\href {\doibase
  10.1016/j.ppnp.2016.01.001} {\bibfield  {journal} {\bibinfo  {journal} {Prog.
  Part. Nucl. Phys.}\ }\textbf {\bibinfo {volume} {88}},\ \bibinfo {pages} {1}
  (\bibinfo {year} {2016})},\ \Eprint {http://arxiv.org/abs/1511.04050}
  {arXiv:1511.04050 [hep-ph]} \BibitemShut {NoStop}%
\bibitem [{\citenamefont {Koch}\ \emph {et~al.}(2017)\citenamefont {Koch},
  \citenamefont {Schlichting}, \citenamefont {Skokov}, \citenamefont
  {Sorensen}, \citenamefont {Thomas}, \citenamefont {Voloshin}, \citenamefont
  {Wang},\ and\ \citenamefont {Yee}}]{Skokov:2016yrj}%
  \BibitemOpen
  \bibfield  {author} {\bibinfo {author} {\bibfnamefont {V.}~\bibnamefont
  {Koch}}, \bibinfo {author} {\bibfnamefont {S.}~\bibnamefont {Schlichting}},
  \bibinfo {author} {\bibfnamefont {V.}~\bibnamefont {Skokov}}, \bibinfo
  {author} {\bibfnamefont {P.}~\bibnamefont {Sorensen}}, \bibinfo {author}
  {\bibfnamefont {J.}~\bibnamefont {Thomas}}, \bibinfo {author} {\bibfnamefont
  {S.}~\bibnamefont {Voloshin}}, \bibinfo {author} {\bibfnamefont
  {G.}~\bibnamefont {Wang}}, \ and\ \bibinfo {author} {\bibfnamefont {H.-U.}\
  \bibnamefont {Yee}},\ }\href {\doibase 10.1088/1674-1137/41/7/072001}
  {\bibfield  {journal} {\bibinfo  {journal} {Chin. Phys.}\ }\textbf {\bibinfo
  {volume} {C41}},\ \bibinfo {pages} {072001} (\bibinfo {year} {2017})},\
  \Eprint {http://arxiv.org/abs/1608.00982} {arXiv:1608.00982 [nucl-th]}
  \BibitemShut {NoStop}%
\bibitem [{\citenamefont {Son}\ and\ \citenamefont
  {Surowka}(2009)}]{Son:2009tf}%
  \BibitemOpen
  \bibfield  {author} {\bibinfo {author} {\bibfnamefont {D.~T.}\ \bibnamefont
  {Son}}\ and\ \bibinfo {author} {\bibfnamefont {P.}~\bibnamefont {Surowka}},\
  }\href {\doibase 10.1103/PhysRevLett.103.191601} {\bibfield  {journal}
  {\bibinfo  {journal} {Phys. Rev. Lett.}\ }\textbf {\bibinfo {volume} {103}},\
  \bibinfo {pages} {191601} (\bibinfo {year} {2009})},\ \Eprint
  {http://arxiv.org/abs/0906.5044} {arXiv:0906.5044 [hep-th]} \BibitemShut
  {NoStop}%
\bibitem [{\citenamefont {Yee}(2009)}]{Yee:2009vw}%
  \BibitemOpen
  \bibfield  {author} {\bibinfo {author} {\bibfnamefont {H.-U.}\ \bibnamefont
  {Yee}},\ }\href {\doibase 10.1088/1126-6708/2009/11/085} {\bibfield
  {journal} {\bibinfo  {journal} {JHEP}\ }\textbf {\bibinfo {volume} {11}},\
  \bibinfo {pages} {085} (\bibinfo {year} {2009})},\ \Eprint
  {http://arxiv.org/abs/0908.4189} {arXiv:0908.4189 [hep-th]} \BibitemShut
  {NoStop}%
\bibitem [{\citenamefont {Son}\ and\ \citenamefont
  {Yamamoto}(2012)}]{Son:2012wh}%
  \BibitemOpen
  \bibfield  {author} {\bibinfo {author} {\bibfnamefont {D.~T.}\ \bibnamefont
  {Son}}\ and\ \bibinfo {author} {\bibfnamefont {N.}~\bibnamefont {Yamamoto}},\
  }\href {\doibase 10.1103/PhysRevLett.109.181602} {\bibfield  {journal}
  {\bibinfo  {journal} {Phys. Rev. Lett.}\ }\textbf {\bibinfo {volume} {109}},\
  \bibinfo {pages} {181602} (\bibinfo {year} {2012})},\ \Eprint
  {http://arxiv.org/abs/1203.2697} {arXiv:1203.2697 [cond-mat.mes-hall]}
  \BibitemShut {NoStop}%
\bibitem [{\citenamefont {Stephanov}\ and\ \citenamefont
  {Yin}(2012)}]{Stephanov:2012ki}%
  \BibitemOpen
  \bibfield  {author} {\bibinfo {author} {\bibfnamefont {M.~A.}\ \bibnamefont
  {Stephanov}}\ and\ \bibinfo {author} {\bibfnamefont {Y.}~\bibnamefont
  {Yin}},\ }\href {\doibase 10.1103/PhysRevLett.109.162001} {\bibfield
  {journal} {\bibinfo  {journal} {Phys. Rev. Lett.}\ }\textbf {\bibinfo
  {volume} {109}},\ \bibinfo {pages} {162001} (\bibinfo {year} {2012})},\
  \Eprint {http://arxiv.org/abs/1207.0747} {arXiv:1207.0747 [hep-th]}
  \BibitemShut {NoStop}%
\bibitem [{\citenamefont {Chen}\ \emph {et~al.}(2014)\citenamefont {Chen},
  \citenamefont {Pang}, \citenamefont {Pu},\ and\ \citenamefont
  {Wang}}]{Chen:2013iga}%
  \BibitemOpen
  \bibfield  {author} {\bibinfo {author} {\bibfnamefont {J.-W.}\ \bibnamefont
  {Chen}}, \bibinfo {author} {\bibfnamefont {J.-y.}\ \bibnamefont {Pang}},
  \bibinfo {author} {\bibfnamefont {S.}~\bibnamefont {Pu}}, \ and\ \bibinfo
  {author} {\bibfnamefont {Q.}~\bibnamefont {Wang}},\ }\href {\doibase
  10.1103/PhysRevD.89.094003} {\bibfield  {journal} {\bibinfo  {journal} {Phys.
  Rev.}\ }\textbf {\bibinfo {volume} {D89}},\ \bibinfo {pages} {094003}
  (\bibinfo {year} {2014})},\ \Eprint {http://arxiv.org/abs/1312.2032}
  {arXiv:1312.2032 [hep-th]} \BibitemShut {NoStop}%
\bibitem [{\citenamefont {Müller}\ \emph {et~al.}(2016)\citenamefont
  {Müller}, \citenamefont {Schlichting},\ and\ \citenamefont
  {Sharma}}]{Mueller:2016ven}%
  \BibitemOpen
  \bibfield  {author} {\bibinfo {author} {\bibfnamefont {N.}~\bibnamefont
  {Müller}}, \bibinfo {author} {\bibfnamefont {S.}~\bibnamefont
  {Schlichting}}, \ and\ \bibinfo {author} {\bibfnamefont {S.}~\bibnamefont
  {Sharma}},\ }\href {\doibase 10.1103/PhysRevLett.117.142301} {\bibfield
  {journal} {\bibinfo  {journal} {Phys. Rev. Lett.}\ }\textbf {\bibinfo
  {volume} {117}},\ \bibinfo {pages} {142301} (\bibinfo {year} {2016})},\
  \Eprint {http://arxiv.org/abs/1606.00342} {arXiv:1606.00342 [hep-ph]}
  \BibitemShut {NoStop}%
\bibitem [{\citenamefont {Tuchin}(2018)}]{Tuchin:2018rrw}%
  \BibitemOpen
  \bibfield  {author} {\bibinfo {author} {\bibfnamefont {K.}~\bibnamefont
  {Tuchin}},\ }\href {\doibase 10.1103/PhysRevC.97.064914} {\bibfield
  {journal} {\bibinfo  {journal} {Phys. Rev.}\ }\textbf {\bibinfo {volume}
  {C97}},\ \bibinfo {pages} {064914} (\bibinfo {year} {2018})},\ \Eprint
  {http://arxiv.org/abs/1802.09629} {arXiv:1802.09629 [hep-ph]} \BibitemShut
  {NoStop}%
\end{thebibliography}%

\end{document}